# An Approach to Twinning and Mining Collaborative Network of Construction Projects

Jia-Rui Lin[1,2,*], Da-Peng Wu[1,3]

[1]*Department of Civil Engineering, Tsinghua University, Beijing, China 100084*

[2]*Tsinghua University-Glodon Joint Research Centre for Building Information Model (RCBIM), Tsinghua University, Beijing, China 100084*

[3]*Beijing YunJianXin Technology Co. Ltd., Beijing, China 100193*

**Abstract**

Understanding complex collaboration processes is essential for the success of construction projects. However, there is still a lack of efficient methods for timely collection and analysis of collaborative networks. Therefore, an integrated framework consisting three parts, namely, system updating for data collection, data preprocessing, and social network analysis, is proposed for the twinning and mining collaborative network of a construction project. First, a system updating strategy for automatic data collection is introduced. Centrality measures are then utilized to identify key players, including hubs and brokers. Meanwhile, information sharing frequency (ISF) and association rule mining are introduced to discover collaborative patterns, that is, frequently collaborating users (FCUs) and associations between information flows and task levels. Finally, the proposed framework is validated and demonstrated in a large-scale project. The results show that key players, FCUs, and associations between information flows and task levels were successfully discovered, providing a deep understanding of collaboration and communication for decision-making processes. This research contributes to the body of knowledge by: 1) introducing ISF and Apriori-based association mining algorithm to identify FCUs and information flow patterns in collaboration; 2) establishing a new data-driven framework to map and analyze fine-grained collaborative networks automatically. It is also shown that people tend to form small groups to handle certain levels or types of tasks more efficiently. Other researchers and industrial practitioners may use this work as a foundation to further improve the efficiency of collaboration and communication.

**Keywords**: social network analysis; digital twin; graph visualization; frequently collaborating users; association rules; process mining; information flow patterns

* Corresponding author: Jia-Rui Lin, Ph.D, Assistant Professor, lin611@tsinghua.edu.cn, jiarui_lin@foxmail.com, Department of Civil Engineering, Tsinghua University

# 1. Introduction

Construction projects are usually complex with intensive labor input, large investment, and extensive involvement of multiple disciplines[1]. One of the paramount challenges currently facing engineers, contractors, and owners is the need to deliver high-quality construction projects in a timely manner. Accurate and timely information of a construction project is needed to achieve well-maintained and efficient project control that will ensure cost and time efficiency of the project[2]. Because 50% to 80% of construction problems arise from a lack of data or a delay in the receipt of information[3], it is essential to collect and analyze data dynamically and make decisions more efficiently. For example, traditional methods for on-site inspection still depend on paper-based files such as drawing and data collection forms, which are not efficient for information exchange and communication, usually leading to overlooked important issues and deferred on-site decisions[4]. Therefore, the success of a construction project calls for fluent and efficient information exchange and collaboration among different stakeholders, including owners, architects, engineers, and contractors[5,6]. Furthermore, timely data retrieval, analysis, and communication of the data in a well-interpreted way are important for construction firms[7,8].

Recently, several efforts have been made to enhance data collection, sensing, and visualization for construction management and collaboration[2]. Various data acquisition technologies, such as Global Positioning System (GPS), Radio-frequency Identification (RFID), barcodes, laser scanners, video and audio technologies for automated data acquisition[9], as well as computer vision and 3D reconstruction[10], have been investigated by both researchers and industrial practitioners. Meanwhile, with the increasing availability of commercial mobile and wearable computers, inspection data collection and knowledge sharing based on mobile devices have also been reported[11,12].

However, there is a lack of research investigating how people collaborate and communicate with each other in a construction project. To the best of our knowledge, most of the relevant studies examined collaboration between stakeholders from an organization-to-organization perspective. For example, a social network method called structural equation modeling was used to analyze the impacts of collaborative relationships on the innovation performance of construction projects[13]. Another study examined how the macro structure of the project-based collaborative network for building information modeling (BIM) implementation in the regional construction industry and public-private sectors evolves over time[14,15]. Usually, a simple graph, that is, an undirected, unweighted graph without multiple edges between two nodes, is built and analyzed to identify potential collaborative patterns[15-18]. In this way, an analysis of the connection strength or communication frequency between two actors, which are usually modeled as multiple edges or weighted edges, is not possible.

Although collaborative networks from a person-to-person perspective have recently been discussed[6], the information flow patterns and frequency of communication are still not considered[19,20]. Nevertheless, investigations have shown that they are important for analyzing and assessing communication and collaboration performance[21]. Thus, the lack of detailed data for creating a fine-grained collaborative network is still an important issue hindering data-driven collaboration and communication. Moreover, it is found that most of the related works establish a collaborative network based on questionnaires or interviews[15,19], which are time-consuming and tedious, making it difficult to obtain timely understanding of collaborative relationships. However, recent attempts to automatically create social networks from log files[6] still suffer from uncertainty, because a number of assumptions are needed to identify nodes and their relationships. That is, an automatic approach to twinning fine-grained collaborative networks is still sought by both researchers and industrial practitioners.

To this end, a novel framework is proposed for: 1) twinning a fine-grained collaborative network of a construction project, 2) detecting key players involved in collaboration, and 3) discovering frequently collaborating users and associations between information flows and task levels. In the proposed framework, a mobile-based application for construction management is extended to collect data related to information flow and collaboration. As a result, a digital twin of the collaborative network could be created. Finally, the social network analysis (SNA) method, including centrality measures, graph visualization, and association rule mining are conducted for knowledge discovery.

In this research, the feasibility of the proposed methodology is first validated and examined. The main objectives of this research are to determine: 1) how to create a digital twin of the fine-grained collaborative network of a construction project, 2) who the key players involved in collaboration are, and 3) how they collaborate and communicate with each other in handling certain construction tasks. In this way, this research contributes to the body of knowledge an innovative framework for twinning and mining detailed collaborative networks of construction projects, and provides insights for project managers by identifying key players and uncovering hidden collaboration patterns from collaborative networks. The remainder of this paper is organized as follows. Section 2 provides a brief review of relevant research and applications. Section 3 introduces the proposed framework and research methodologies, including statistical analysis and SNA. In Section 4, a demonstration of the proposed method as well as discovered key players, frequently collaborating users, and associations between information flows and task levels are presented. Section 5 discusses the findings and limitations of study. Finally, the paper is concluded in Section 6 with a summary of the contributions to the construction domain and a discussion of possible future improvements.

## 2. Overview of Related Research

Construction projects are usually complex and exhibit highly fragmented operation, with intensive labor input, large investment, and extensive involvement of multiple disciplines[1]. Several studies have shown that a construction project failure is closely associated with communication between stakeholders[6]. On one hand, 50% to 80% of construction problems arise from a lack of data or a delay in the receipt of information[3], whereas accurate and timely shared information of a construction project is needed to ensure its cost and time efficiency[2]. On the other hand, successful innovation usually requires effective cooperation and working relationships among different parties within a construction project[13]. Thus, there was a call for closer relationships between clients, designers and contractors, which requires new tools to create and maintain collaborative relationships between multiple stakeholders through effective information exchange and communication patterns[22].

### 2.1 Collaborative Relationships

Collaboration is a process in which individuals or organizations work together and they typically establish collaborative relationships to share information and resources, thereby increasing mutual benefits compare to working alone[13]. Generally, individuals or organizations are modeled as nodes, and their connections or relationships are considered as edges or ties, thereby forming a social network[23]. Collaborative relationships reveal how actors (people, resources, etc.) interact with each other; they are widely used in analyzing the innovation of organizations[13], safety climate[24] and communication[19], gaps for project success[22], collaboration barriers[17] in the construction domain and in other areas such as research collaboration[25], collaborative learning[26], business[27] and software development[28] (Table 1).

Researchers typically investigate collaborative relationships from two levels: the network level and node level. The former considers collaboration through the overall characteristics of a network. For example, a BIM-based collaborative environment for the design of green buildings was proposed[29], in which network-level metrics such as network density and diameter are calculated for socio-technical analytics. The latter pays more attention to key players in a network; for example, central actors were identified from a project-level social network for successful project delivery[6]. Another study examined how collaborative relationships impact the innovation performance of construction projects, and key members in the collaborative networks were identified[13]. Frequently, both network-level and node-level features are considered to investigate collaborative relationships. With a longitudinal data set of projects in China, the overall characteristics and individual performance of the collaborative network were analyzed and it was illustrated that the collaborative network became increasingly dense around certain key nodes[14]. From an organization-to-organization perspective, the formation of a construction firm's collaborative network for performing international projects was investigated, and it was concluded that large companies and small and medium-sized companies have different tendencies when pursuing collaborative ventures for overseas construction projects[17]. Meanwhile, an examination of project-level networks showed low levels of network density in general and high centrality for certain actors, reflecting their dominant role in construction projects[22].

### 2.2 Social Network Analysis Method

Because most of the related works view the problem from a network perspective, the SNA method is usually adopted to investigate features of collaborative networks[22]. SNA enables systematic specification of the relationships between actors within a group, and the results can be represented mathematically and graphically[18].

To examine the overall characteristics, network-level metrics such as network size, density, diameter, and average node degree are usually adopted[15]. Among them, network density is the most commonly used (Table 1), which describes the portion of potential connections in a network that are actual connections[20]. In this manner, network density is usually adopted to quantify how tightly different actors are connected.

At the same time, several centrality measures (Table 1) including degree of node or degree centrality (DC), betweenness centrality (BC), closeness centrality (CC) and eigenvector centrality (EC) are introduced to quantify the importance or influence of a particular node in SNA[15]. Thus, centrality measures provide a way to identify key players at a micro-level. The first and simplest centrality introduced is DC, which is the total number of edges directed to a node and edges directed from the node to others. In a collaborative network, DC indicates the popularity of a node. Apart from DC, CC and BC are two commonly utilized centrality measures. The CC of a node is the average length of the shortest paths between the node and all other nodes in the graph, which quantifies how close the node is to others[30]. BC is the number of shortest paths that pass through the node, and quantifies the number of times the node acts as a bridge along the shortest path between two other nodes[31]. Thus, DC and CC may be used to identify potential hubs or actors[13] in a social network, and BC helps to detect brokers in a collaborative network[6]. The calculation methods of degree centrality $D(x)$, closeness centrality $C(x)$, and betweenness centrality $B(x)$ are, respectively, represented as follows:

$$D(x) = Count(E_k | StartNode(E_k) = x) + Count(E_k | EndNode(E_k) = x) \qquad (1)$$

$$C(x) = 1 \Big/ \sum_{y} d(y, x) \tag{2}$$

$$B(x) = \sum_{y \neq x \neq z} \sigma_{yz}(x) / \sigma_{yz} \tag{3}$$

where $x, y, z$ are different nodes, $d(y, x)$ is the distance between nodes $x$ and $y$, $\sigma_{yz}$ is the total number of shortest paths from node $y$ to node $z$, and $\sigma_{yz}(x)$ is the number of those paths that pass through node $x$.

Moreover, community detection is also adopted to discover small subgroups in a large and dense network[6]. Thus, it is possible to capture the spontaneous formal and informal interaction between project actors at different project levels[22] and develop collaboration strategies to achieve better outcomes while considering the relevant network patterns[17]. Consequently, by analyzing both stakeholders and their interests from a network perspective, SNA can improve the accuracy, completeness and effectiveness of stakeholder management in construction[32].

Although the above-mentioned methods are popular for investigating collaborative patterns, they mainly focus on the overall features or key players of a network and neglect the differences between the connections of different nodes. This is because the adopted metrics take all connections with the same weight. In other words, evaluating how tightly all individuals are connected and how important an actor is in a collaborative network is possible, whereas it is still impossible to tell whether one connection is more important than another or is there any association between them.

## 2.3 Data Collection and Characteristics of Network

In contrast to building social networks based on academic databases for research collaboration[25], surveys and interviews are two commonly used methods[15,19] to create collaborative networks in the construction domain, which are time-consuming, tedious and error-prone. Another method introduced recently is to create social networks based on log files generated by software, for example, for collaborative learning[26] and design[6]. When adopting this method, several assumptions are needed to identify nodes and their connections, which may introduce uncertainty for network creation. Most importantly, details of the collaborative network are always missed with these methods and timely analysis of the network is somewhat challenging. This is why the networks created in previous works are typically simple, unweighted and undirected graphs, with the number of nodes and ties or edges between 10 and 1, 000 (Table 1). For example, most of the related works[15-18] only built an organization-to-organization social network without considering the weight of edges, which represents the connection strength between different firms, or how closely two companies collaborate with each other. Although person-to-person collaboration or communication networks are presented in certain studies[19,20], the direction and frequency of information flow between two individuals, that is, direction and weight of network edges, are still not considered, which are important for analyzing and assessing the performance of communication and collaboration[21].

Owing to a lack of detailed data, it is still impossible to analyze more detailed collaborative patterns such frequently collaborating users (or strong connections between nodes) and information flow patterns. Moreover, the lack of automatic methods to create collaborative networks in a timely manner also hinders efficient decision-making process. Therefore, a method that automatically collects collaboration data and builds collaborative networks dynamically would be interesting to both researchers and practitioners to deepen their understanding of collaboration.

Table 1 Summary of Related Literatures

| Area | Contribution | Data Collection | Metrics Used* | Graph Type** | No. of Nodes | No. of Ties |
|---|---|---|---|---|---|---|
| Academe | Research collaboration and productivity[25] | Academic Database | Network: Density<br>Centrality: B, C, D, E<br>No. of Links | Simple | 708 | 847 |
| Business | Analyzing collaborative patterns[27] | Survey | Network: Size, Density<br>Centrality: B, C, D | Simple | 24 | N/A |
| Learning | Analyzing communication patterns and interaction[33] | Interview | Centrality: B, C, D, E | Simple | 19 | 60 |
| Software | Collaboration patterns and impact on awareness[28] | Interview<br>Survey | Communication Frequency | Simple | 101 | 140 |
| Construction | Comparing knowledge integration in competitive and collaborative working[20] | Interview | Network: Density<br>Centrality: D | Simple | 9 | 26 |
| Construction | Analyzing project coalition[18] | N/A | Centrality | Simple Directed | 20 | 43 |
| Construction | Network gaps and project success[22] | N/A | Network: Density<br>Centrality: D | Simple Directed | N/A | N/A |
| Construction | Analyzing collaborative venture[17] | Government Data | Network Density<br>Centrality: B, C, D<br>No. of Triad | Simple | 133 | 389 |
| Construction | Collaborative design[29] | N/A | Network: Density, Diameter<br>Network Size | N/A | N/A | N/A |
| Construction | Collaborative innovation[13] | Interview | Network: Density<br>Centrality: B, C, D | Simple | 16 | 135 |
| Construction | Longitudinal analysis of macro-level collaboration[15] | Interview | Network: Density, Distance<br>Average Node Degree<br>Clustering Coefficient | Simple | 212 | 747 |
| Construction | Safety communication[19] | Survey | Network: Density<br>Centrality: D | Simple Weighted | 183 | N/A |
| Construction | Collaborative design[6] | Log Files | Network: Density<br>Centrality: B, C, D | Simple | 23 | 218 |
| Construction | Collaborative patterns and power imbalance[16] | Interview | Network: Density<br>Centrality: B, C, D | Simple | 41 | 163 |
| Construction | Communication assessment of change management process[21] | Unknown | Network: Density, Diameter<br>Average Node Degree<br>Average Strength<br>Centrality: B, C, D | Weighted Directed | 412 | 3402 |
| Learning | Interaction pattern in networked learning[26] | Log Files | Network: Density<br>Centrality: Out-D | Simple | 8 | 61 |
| **Construction (Current)** | **Network twinning and mining of collaborative patterns** | **Automatic** | **Network: Density, Diameter<br>Centrality: B, C, D<br>Information Sharing Frequency<br>Association Rules** | **Directed with Multiple Edges** | **226** | **17068** |

* Centrality: B-Betweenness, C-Closeness, D-Degree, E-Eigenvector   ** Simple denotes a unweighted, undirected graph without multiple edges between two nodes

# 3. Methodology

Construction projects consist of various tasks, which involve many stakeholders and information transfers. A digital twin of a fine-grained collaborative work could help managers and decision-makers to identify: 1) the key players involved in the collaboration, 2) any frequently collaborating users, and 3) any associations between information flows and types or severity levels of tasks.

To twin a collaborative network of a construction project automatically and discover hidden knowledge for better decision-making, this research proposes an integrated approach to dynamically collect and quantitatively analyze collaboration data of construction projects. As illustrated in Figure 1, a research framework consisting of three procedures is adopted: 1) automatic data collection by updating existing systems with specific components, which could collect detailed data related to collaboration and store them in a database automatically; 2) data preprocessing to clean collected raw data and build collaborative networks; 3) identification of key players (hubs and brokers) and frequently collaborating users, and discovering association rules between information flows and task levels, thus providing deep insights on how people collaborate in a construction project and improve the decision-making process. To implement the proposed framework, WeChat-based Mobile App for data collection, PostgreSQL for data persistence, python-based scripts, and tools such as NetworkX, Gephi are adopted. Finally, the proposed method is applied and tested in a realistic construction project to validate and improve the proposed approach in this research. Each of these procedures is discussed below.

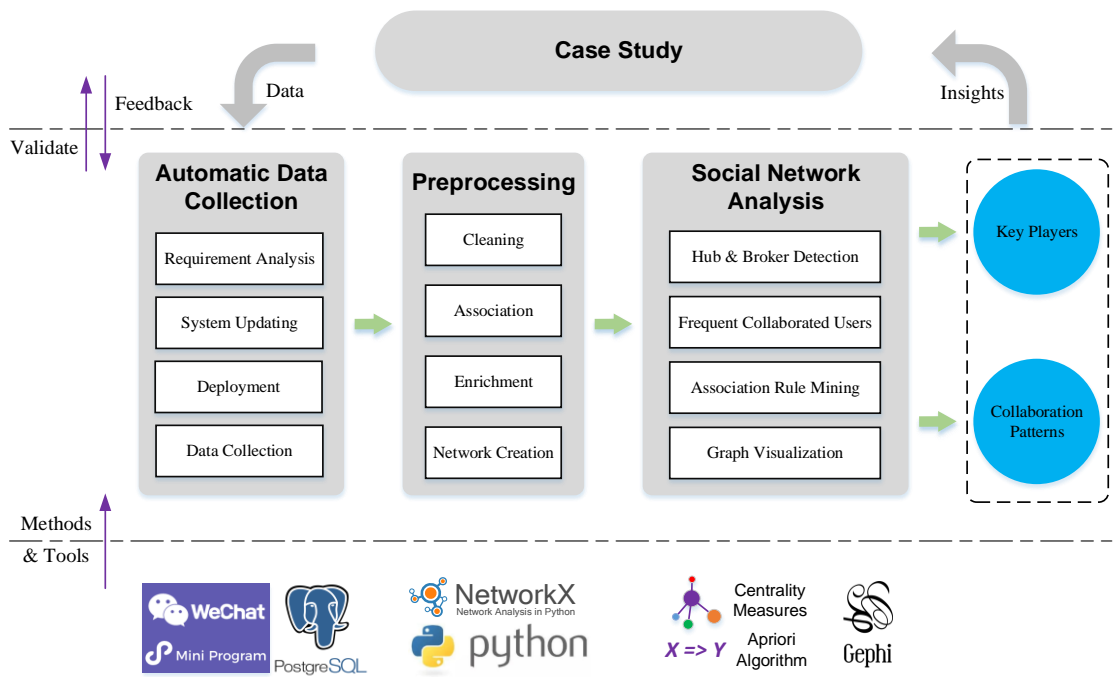

Figure 1 Flowchart of the Proposed Methodology

## 3.1 Automatic Data Collection for Collaborative Network Twinning

Nowadays, there are already many information systems utilized in construction projects for various tasks, such as safety management systems for bridge construction[1] and information-sharing platform for multiple stakeholders[5]. However, most of them focus on tracking, evaluating the status of constructed facilities and improving data interoperability among different systems, and they lack specific components to collect collaborative data of the users of these information systems. Therefore, updating an existing system is recommended to enable the twinning capacity of collaborative networks.

To achieve automatic data collection, information requirements should initially be determined. That is, what types of data are needed, what are the granularities of the data, and how frequently are they collected? Then, according to the extracted information requirements, the development of new components or extension of existing systems, which is called the system updating strategy in this research, is recommended to automatically collect the required data. Meanwhile, collecting these data in the background such as using a software-based logging module is suggested. In this way, it is possible to collect timely communication and collaboration data without any interrupting to the normal workflow. Finally, the updated system can be deployed and utilized for daily communication and collaboration, as usual, and the data will be collected and stored in a database automatically.

In this research, a mobile app called MobileCM for on-site inspection and construction management[34] was used as a starting point. Because a fine-grained collaborative network is needed in this research, the team decided to collect the time, creator, sender, and receiver of each on-site inspection issue. Thus, along with the developed functions for on-site safety and quality inspection, a small component is injected into MobileCM to collect data related to collaboration. Specifically, for each inspected on-site issue, an issue record is created as normal, and the creator and time when the issue record is created and shared with others is recorded. Meanwhile, the sender and receiver of an issue record are also captured when an engineer or construction manager shares it with others. With the upgraded MobileCM, engineers and other users simply use it as usual and collaboration-related data can be collected timely and automatically. All the collected data are stored in a relational database called PostgreSQL[34]. Figure 2 provides an example of how issue records, data transfers (or issue forwards) are stored in PostgreSQL. In the same way, it is easy to upgrade existing systems and collect collaboration data for other tasks involved in construction projects.

Figure 2 Collected Data Related to Collaboration

To date, the upgraded MobileCM application has been deployed and used in more than 30 projects in China since December, 2018, and the current study selects data of 1 typical project with approximately 7800 issue records from them for research purpose.

## 3.2 Data Preprocessing

As mentioned above and in our previous work[34], the collected data are stored in a relational database called PostgreSQL, and issue records, forward records, and user information are all stored in different data tables. Moreover, the quality of collected data greatly impacts the data mining process conducted later. Therefore, a four-step data preprocessing process is adopted:

1) Data cleaning: issue records whose descriptions are empty or too short (less than 5 words in this research), and those without creation time are omitted. Forward records without a sender or receiver are also skipped. This is achieved by specifying certain constraints when querying the database. For example, the following structured query language (SQL) query skips all issue records with short or empty descriptions:

```
SELECT * FROM issue_record WHERE Description IS NOT NULL AND LENGTH(Description)>5
```

2) Data association: users involved in the project have different roles and belong to different organizations; because these data are kept in different tables, a data association process is needed to connect or join different tables. Similarly, issue records and issue forward records should be associated too. In this research, part of the data association is implemented based on JOIN clause of an SQL query, and the other part is performed using Python scripts.

3) Data enrichment: raw data extracted from the database lacks semantic information for analytics purposes. In this research, the meanings of TypeID, LevelID, RoleID, etc. are added according to discussions with software developers and product managers.

4) Network creation: to handle a certain inspected on-site issue (or a construction task), submitted issue records are usually forwarded from one user to another; that is, a small network could be formed by combining the issue record and corresponding issue forward records as well as data of involved users. For example, the upper-left corner of Figure 3 shows that an engineer $U_1$ inspected an on-site issue and submitted an issue record to his/her manager $U_2$; then, the issue is assigned to $U_3$ for further resolution and feedback is returned to $U_2$. Finally, the results are submitted to $U_4$ for a final check. In this way, the handling process for each on-site issue or task is represented as a small network, and combining all the small networks creates a large collaborative network of the construction project. The right part of Figure 1 illustrates how the network is created. First, we iterate all the issue records and filter related issue forward records for each issue record, then find or create network nodes based on the source user ID and target ID of each issue forward record, and finally create an edge between the two nodes with the ID and creation time of the issue forward record. It should be noted that the type, severity level, and other fields of an issue record are also attached to network edges as properties. Similarly, the roles and organizations of users are also attached to network nodes. Consequently, a multiple-directional graph or collaborative network is established.

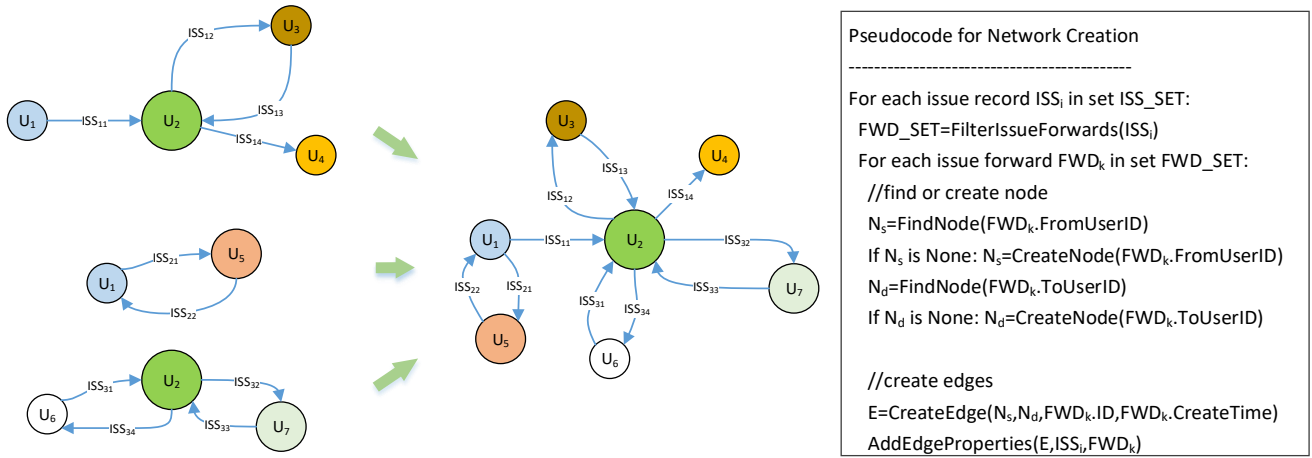

Figure 3 Examples and Pseudocode for Network Creation

In this step, NetworkX, a widely used Python package for network creation and analysis, as well as other Python-based packages are used. With a Python-based database driver, SQL queries used for data cleaning and association are executed and the cleaned data is retrieved from the PostgreSQL database. Then, data enrichment is conducted by iterating each retrieved data record and attaching required semantic information as mentioned in step 3). Finally, the collaborative network is created through a NetworkX-based implementation of the pseudocode shown in Figure 3. The generated network is a multiple-directed graph, that is, graph with multiple directed edges between different nodes. Moreover, the generated network is also exported as a gml file based on NetworkX, and could be further used for network visualization.

## 3.3 Social Network Analysis

Because people usually assume different roles in a collaborative network, the identification of different roles provides managers with a thorough comprehension of the social system of construction projects[6]. Moreover, there are many issue records generated and forwarded from one stakeholder to another every day. Detection of collaborative patterns (e.g., frequently collaborating users) helps decision-makers understand how people collaborate and optimize the social system of construction projects. The following sections explain the measures and methods used for hub/broker detection and identification of collaborative patterns, more specifically, frequently collaborating users and associations between information flows and task levels.

(1) Hub and Broker Detection

In this research, centrality measures including BC, CC, and DC are used to identify users who act as a hub or a broker in sharing information or resolving on-site issues. With identified key players, managers could improve the efficiency of the collaboration and decision-making process.

DC and CC are usually utilized to find hubs in a network, because they quantify how frequently a user is connected with others and how easily a user can share information with others respectively. Thus, a higher DC of a user means he/she receives or sends more information from or to others, which implies that he/she is in a position for on-site inspection or issue resolution. Similarly, a higher CC implies that a user has a greater ability of data dissemination and reception, and he/she is likely a manager or coordinator in the collaborative network.

Another important role is that of the broker, who works as a bridge to connect different engineers or workers. To detect brokers in a collaborative network, BC is adopted. As mentioned before, BC counts the number of shortest paths passing through a node, and quantifies the efficiency of a node in information sharing. The difference between CC and BC is that users with higher CC have more shortest paths to others, that is, they can spread information to others more easily, whereas individuals with higher BC have a greater ability to transfer information from one to another, which means that their absence would decrease communication efficiency dramatically.

In this research, a graph mining and visualization tool called Gephi[35] is utilized, owing to its built-in support for the calculation of various centrality measures, including BC, CC, DC, and even EC. The gml file generated in the previous step can be imported into Gephi directly, and the required centrality measures of a collaborative network can be obtained. Moreover, Gephi also provides an intuitive view of the collaborative network with various visualization features.

(2) Frequently Collaborating Users

Because information is transferred between different users, it is quite common for the number of data transfers between a certain pair of users to be much higher than those between the others. In this research, information sharing frequency (ISF) is introduced to identify frequently collaborating partners in a quantitative way. The calculation of ISF is as follows:

$$isf(x,y) = Count(E(x,y)) \geq ISF_{min} \qquad (4)$$

That is, the ISF between nodes $x$ and $y$ can be calculated by counting the number of edges directly connecting the two nodes. Furthermore, given that on-site issues or other construction tasks are categorized into different groups based on levels or other labels, a fine-grained measure called labeled information sharing frequency (LISF) can be defined as:

$$lisf(l_i, x, y) = Count(E(x,y)|l_i) \geq LISF_{min} \qquad (5)$$

That is, LISF is calculated by iterating all edges connecting nodes $x$ and $y$ and counting the edge with the label equals to $l_i$.

Then, it is easy to find node pairs with the highest ISF or LISF, which indicates that the identified user pairs are the most active and important in handling construction tasks or a certain group of thereof. It is also possible to identify node pairs with minimum thresholds $ISF_{min}$ or $LISF_{min}$. Paying more attention to them could help improve the overall work efficiency for construction management.

As mentioned earlier, the generated collaborative network is a multiple-directed graph and there are multiple directed edges between users. With a defined ISF, a simple graph is generated by merging multiple edges between nodes and assigning the ISF as the weight of edges. Then, frequently collaborating users could be identified by filtering edges with minimum weights greater than or equal to $ISF_{min}$. Similarly, if we only merge edges with the same label, and filter edges by $LISF_{min}$, frequently collaborating users related to different task levels could be identified. Because Gephi provides rich features for network editing such as edge merging and filtering, it is also used for the identification of frequently collaborating users.

(3) Associations Rules of Information Flow

To further understand how information flows between users in the collaborative network are associated, or if an issue record or task is transferred from user $x$ to $y$, how likely will it be sent from user $y$ to $z$, a frequent itemset mining method is used to find information flow patterns among involved users, namely, association rules between different edges.

Generally, suppose $ET = \{T_1, T_2, ..., T_n\}$ is a set of transactions, and each transaction $T_i = \{E_1, E_2, ..., E_k\}$ is a set of items, then, subsets of a transaction are usually called itemsets. If $EA$ and $EB$ are two itemsets, and $EA \Rightarrow EB$ represents the association between them, the following metrics are defined[36,37]:

$$Support(EA \Rightarrow EB) = \frac{|EA \cup EB|}{|ET|} \times 100\% \geq S_{min} \qquad (6)$$

$$Confidence\left(EA \Rightarrow EB\right) = \frac{Support\left(EA \Rightarrow EB\right)}{Support(EA)} \times 100\% \geq C_{min} \qquad (7)$$

$$Lift\left(EA \Rightarrow EB\right) = \frac{Confidence\left(EA \Rightarrow EB\right)}{Support(EB)} \geq L_{min} \qquad (8)$$

where $|EA \cup EB|$ is the frequency of itemset $EA \cup EB$, or how many transactions has a subset $EA \cup EB$, and $|ET|$ is the total number of transactions. In this way, support quantifies the popularity of an itemset among all transactions, confidence is the likelihood that $EB$ occurs if $EA$ occurs[38], and lift reflects the increase ratio of occurrences of $EB$ when $EA$ occurs. A lift of 1 means there is no association between $EA$ and $EB$, and lift greater than 1 means they will occur together more likely. By choosing proper values of $S_{min}$, $C_{min}$, $L_{min}$, minimal thresholds of the defined metrics respectively, potential associations between different itemsets could be identified.

Specifically, information flows in handling a certain issue record or task could be taken as a transaction $T_i$, and every data forward from user $x$ to $y$ (or edge $E_{xy}$) is an item of $T_i$. Thus, if antecedent $EA$ is $\{E_{xy}\}$ and consequent $EB$ is $\{E_{yz}\}$, the association between $E_{xy}$ and $E_{yz}$, $E_{xy} \Rightarrow E_{yz}$, is the same as $EA \Rightarrow EB$, which could be evaluated based on Equation (6)-(8). Then, how likely the information would flow from user $y$ to $z$ provided that it has flowed from user $x$ to $y$ could be quantified by confidence, and its popularity and increase ratio are also quantitatively described by support and lift. Moreover, considering that both $EA$ and $EB$ are itemsets, or set of edges, association rules between multiple information flows could also be identified.

Meanwhile, tasks with different types or levels are usually handled in different ways, and thus it is interesting to further explore how they will impact information flow and collaboration patterns. Therefore, the label (e.g., level or type of an issue record) of edges is considered to discover association rules between categories of issues (or construction tasks in general) and information flows. The difference is only edges with the same label are considered when calculating support, confidence or lift. Thus, associations between information-sharing paths and task levels or types can be identified.

Based on these equations, the Apriori algorithm is adopted to find association rules between different edges. By iterating the created collaborative network, all edges between different nodes as well as the corresponding properties of related issue records are converted into a list of itemsets through Python scripts. Then, the Python-based Apriori module is adopted to find association rules between different edges considering specified minimum values of the above-defined metrics.

## 4. Case Study

To validate the usefulness of the proposed approach for the twinning and mining of construction project collaborative networks, a realistic project in China was selected. Previously developed MobileCM was upgraded to include the components mentioned in Section 3.1 for collecting data on collaboration during the on-site inspection and resolving process[34]. With the developed MobileCM, on-site issues related to safety and quality are collected, and divided into three different severity levels: low, medium, and high. Meanwhile, different statuses (e.g., submitted, assigned, pending, approved) of submitted issues are also recorded in the database. As for the users, a number of roles including owner, safety engineer, safety manager, and project manager, are defined and assigned to each user record.

As shown in Table 2, a new version of MobileCM was deployed and data from March 20, 2019 to October 28, 2019 were collected. During this period, 7821 issue records and 23517 issue forward records were generated and collected in total. Meanwhile, 709 users were registered in the system. Following the data preprocessing process mentioned in Section 3.2, 7250 issue records, 17068 issue forwards and 226 user records were retained for data mining purposes. Figure 4 shows the number of data records generated monthly. Except for March, which is the beginning of data collection, approximately 800-1200 on-site issues, and 2700-4000 issue forwards were collected every month. Usually, for each issue record, approximately 3~3.7 issue forward records were generated and collected on average.

Table 2 Statistics of Utilized Data

| Data Type | Count | Count After Cleaning | Period of Time |
|---|---|---|---|
| Issue Records | 7821 | 7250 | 2019/03/20-2019/10/28 |
| Issue Forwards | 23517 | 17068 | 2019/03/20-2019/10/28 |
| Users | 709 | 226 | / |

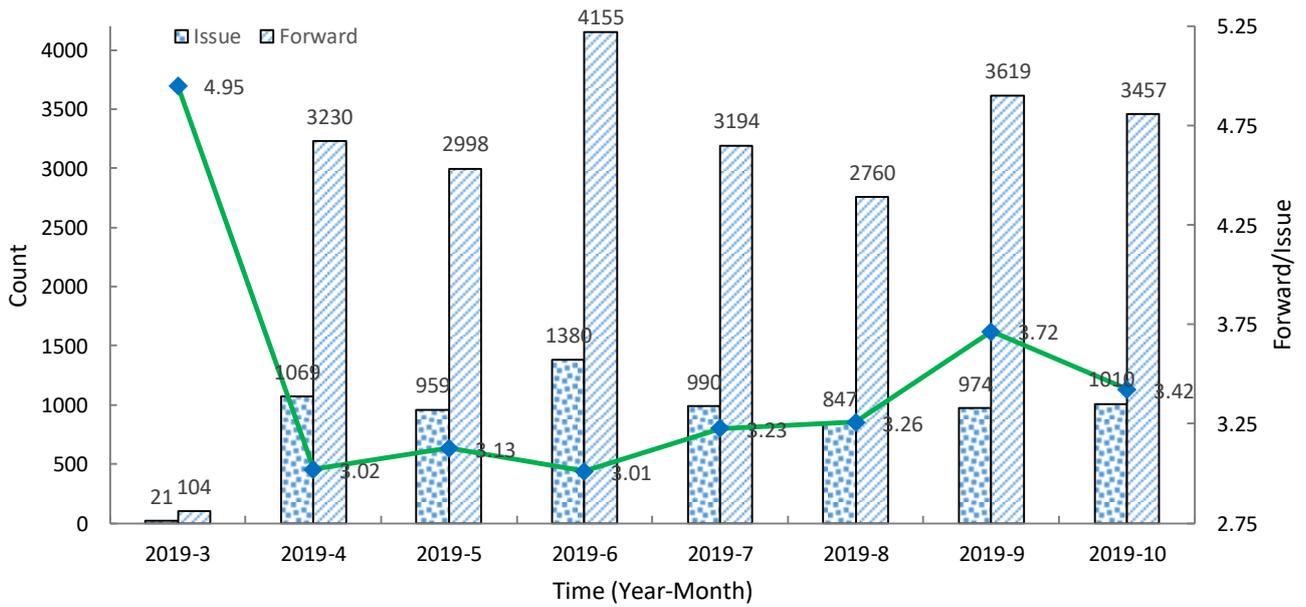

Figure 4 Number of Issues, and Forwards Generated Every Month

Figure 5 shows the generated collaborative network with node colors indicating different user roles and edge colors presenting types of issue records. In total, nine user roles (engineers from the owner's company, project supervisors from the supervision company, and other roles from the general contractor) are involved in the collaboration. It is found that most of the inspected issues are related to safety and the number of issues handled by different users is significantly different.

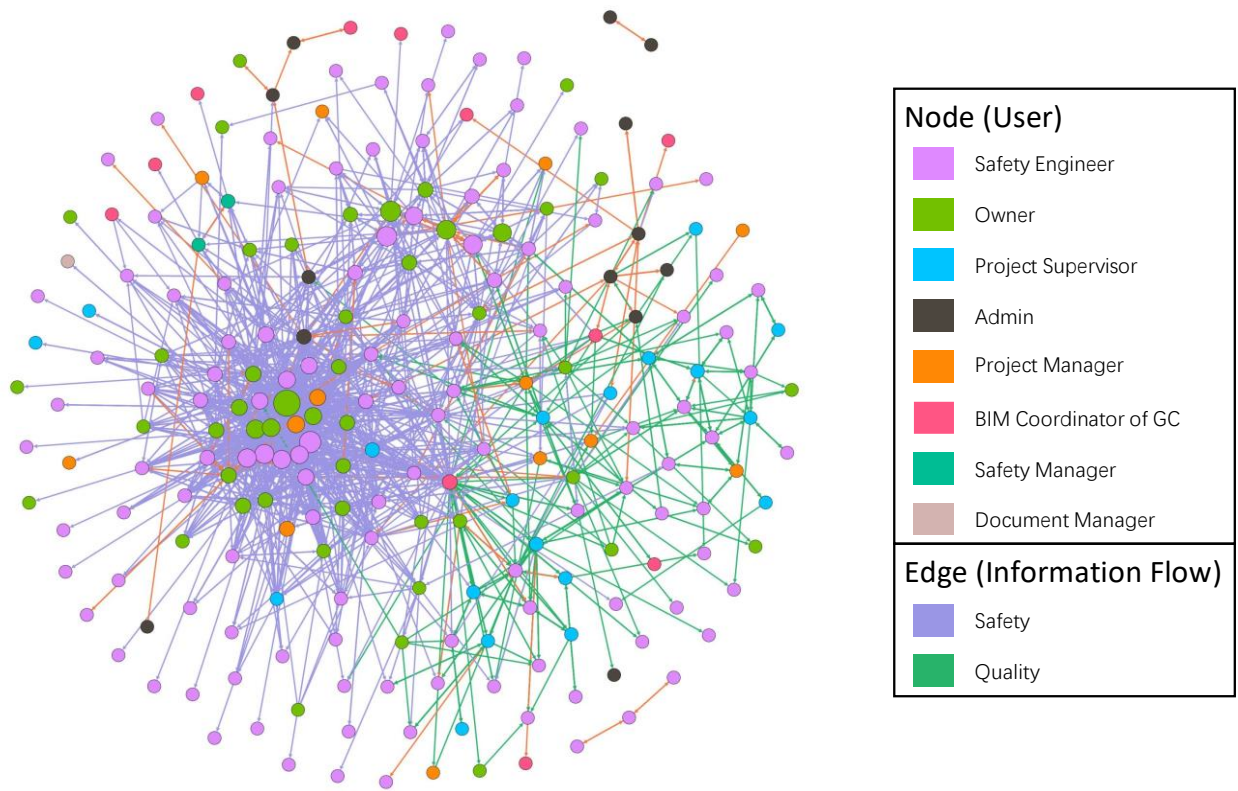

Figure 5 Generated Collaborative Network

As mentioned earlier, 226 users are involved and 17068 issue forwards are created at the same time; thus, a social network with 226 nodes and 17068 edges is obtained. Table 3 shows some basic indicators suggested by previous works[6] to quantify the overall characteristics of the generated collaborative network. For example, the average degree of the network is 75.5, indicating that each involved user in the network is connected to 75-76 other participants on average. Density reflects connectivity of a network, such that the highest value of 1.0 means that all users are connected to each other, and the lowest value of 0.0 means no connection exists among the participants. For the current project, the density of the network is 0.336. That is, 33.6% of all possible links are present, representing a medium level of network cohesion. The diameter of the network quantifies the greatest distance between two nodes in a network, in this case, it is 8. Modularity measures the strength of the division of a network into groups or communities. A network with high modularity usually has dense links between nodes within communities but sparse connections between nodes in different groups. As the range of modularity is between -0.5 and 1.0, a value of 0.489 implies that the network could be divided into a few communities.

Table 3 Basic Features of Generated Collaborative Network Measures

| Attribute | Value | Attribute | Value |
| --- | --- | --- | --- |
| Number of Nodes | 226 | Density of Network | 0.336 |
| Number of Edges | 17068 | Diameter of Network | 8 |
| Average Degree | 75.5 | Modularity of Network | 0.489 |

## 4.1 Detection of Key Players

Given that different participants are highly related to each other with a high average degree, it is important to identify key players in the network. Therefore, centrality measures defined in Equations (1), (2), and (3) are typically used. Generally, DC helps in finding hub users with the highest connections to and from them, that is, worker who handles most of the information flow. CC identifies participants with the lowest average distance to others, and is used to find hub users that are able to spread information very efficiently in the network. On the contrary, BC pays more attention to users who function as connectors or brokers between other users, and quantifies the influence of a node over the information flow. In Table 4, the top 16 nodes in the network based on DC, CC, and BC are provided.

Table 4 Top-16 Nodes based on Different Centrality Measures

| Rank | Degree Centrality | | | Closeness Centrality | | | Betweenness Centrality | | |
| --- | --- | --- | --- | --- | --- | --- | --- | --- | --- |
| | ID | Role | Degree (In+Out) | ID | Role | Closeness | ID | Role | Betweenness |
| 1 | 255 | O | 2371 = 1176 +1195 | 12 | C | 0.5082 | 255 | O | 6962.5 |

| | | | | | | | | | | | |
|---|---|---|---|---|---|---|---|---|---|---|---|
| 2 | 62 | S | 1471 = | 745 | +726 | 62 | S | 0.4331 | 12 | C | 5426.7 |
| 3 | 210 | O | 1276 = | 637 | +639 | 67 | S | 0.4314 | 62 | S | 3052.1 |
| 4 | 75 | S | 1174 = | 585 | +589 | 275 | S | 0.4272 | 149 | S | 2399.9 |
| 5 | 278 | S | 1067 = | 539 | +528 | 262 | S | 0.4197 | 261 | M | 2196.2 |
| 6 | 204 | O | 1040 = | 519 | +521 | 66 | S | 0.4181 | 177 | P | 1961.4 |
| 7 | 231 | S | 1007 = | 506 | +501 | 255 | O | 0.4165 | 210 | O | 1890.5 |
| 8 | 258 | O | 954 = | 477 | +477 | 202 | M | 0.4165 | 65 | S | 1552.3 |
| 9 | 312 | S | 939 = | 471 | +468 | 289 | S | 0.4133 | 289 | S | 1484.8 |
| 10 | 262 | S | 913 = | 464 | +469 | 261 | M | 0.4125 | 75 | S | 1457.7 |
| 11 | 199 | O | 898 = | 343 | +555 | 286 | S | 0.4087 | 198 | O | 1426.2 |
| 12 | 259 | O | 827 = | 417 | +410 | 208 | S | 0.4018 | 51 | O | 1421.5 |
| 13 | 67 | S | 825 = | 415 | +410 | 65 | S | 0.3974 | 282 | S | 1407.0 |
| 14 | 264 | S | 825 = | 409 | +416 | 129 | S | 0.3960 | 275 | S | 1385.1 |
| 15 | 65 | S | 719 = | 370 | +349 | 214 | O | 0.3889 | 315 | P | 1338.1 |
| 16 | 211 | O | 718 = | 349 | +369 | 97 | S | 0.3882 | 285 | S | 1286.2 |

*O-Owner, S-Safety Engineer, C-BIM Coordinator, M-Project Manager, P-Project Supervisor

(1) Analysis of Detected Hubs

According to DC (first four columns of Table 4), user #255 is the top one as she or he sends and receives the most data records, which is 2371 in total. Furthermore, except for user #199, the number of data sent and received by the top 16 participants is essentially the same. This is consistent with the workflow verified by the owner, general contractor, and supervision company, as the owner requires that all issue records and their handling results should go back to the user who submitted them for a double check. Because the dominant type of on-site issue is safety (edges colored in purple in Figure 5 and Figure 6), it is reasonable that the roles of the top 16 hub users based on DC are either owners or safety engineers, which are also illustrated in the left part of Figure 6 as green and purple nodes.

When considering CC (columns 5 to 7 of Table 4), user #12 is selected as the top hub, which is consistent with the role of user #12 as the BIM coordinator, and with their main responsibility being to coordinate the work of others. Similarly, project managers #202 and #261 are also selected as hubs, because their role is to organize all the construction tasks and keep them moving forward fluently. Although user #255 has the highest DC, his or her CC is not that high, indicating that the ability to spread information of user #255 is lower than that of certain other users such as #12, #62, and #67. It is also concluded that user #62 ranks second among all the users for both DC and CC. In other words, user #62 has a great ability to receive information and spread it to others. This is also confirmed by the leaders of this project, as user #62 is usually selected as the first engineer to receive observed on-site issues by the owners (a strong connection between users #62 and #255 is also discovered in Section 4.2). Following user #62, there are also a few safety engineers (users #67, #275, #262, #66, etc.) working as hubs to handle and spread submitted on-site issues. Finally, the results also show that the difference in closeness among the top 16 users is not so large, which means that it is easier to find another user to handle the submitted issues if the current user could not process them in time.

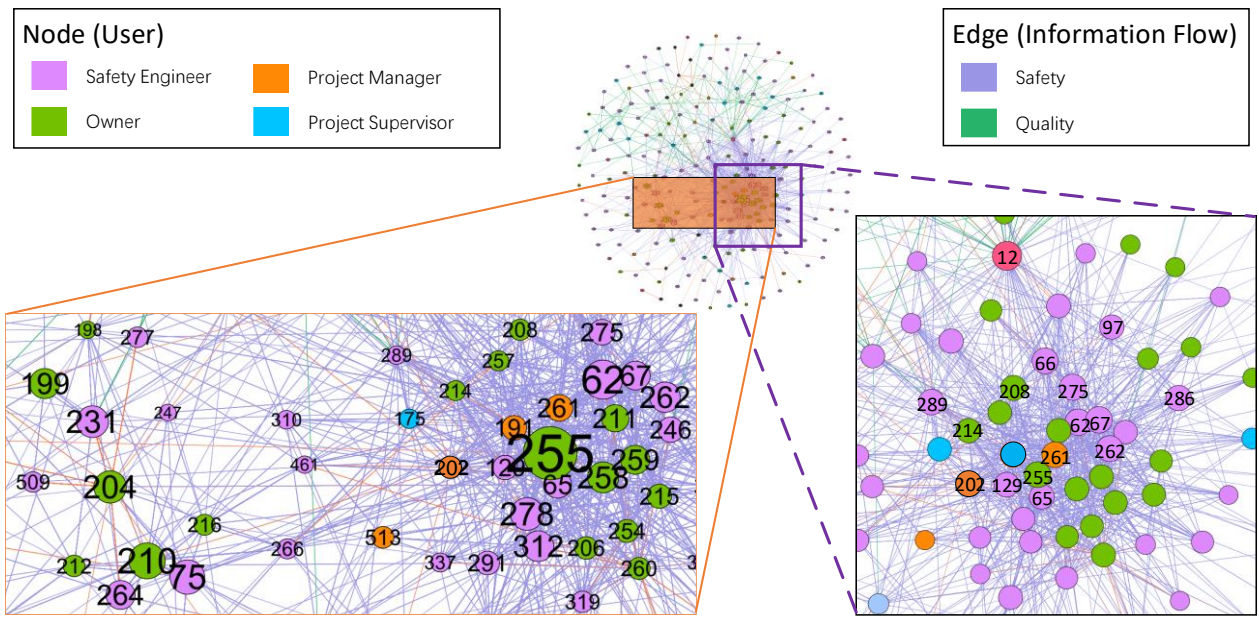

Figure 6 Detected Hubs based on Degree (Left) and Closeness (Right) Centralities

In addition, users #255, #62, #262, #67, and #65 are identified as hubs by both DC and CC, which highlights their importance in handling and spreading information. Moreover, Figure 6 also shows that the top 16 hubs by either DC or CC are all in the central part of the network, as they usually have more connections with others.

(2) Analysis of Detected Brokers

The last three columns of Table 4 list the top 16 users based on BC. It is found that users #255, #12, and #62 are the top three brokers. Their roles are owners, BIM coordinators, and safety engineers, respectively, and they could build connections between different groups. Compared with DC and CC, an additional user role, project supervisor, is also included in the top 16 user list by BC. In their positions as project supervisors, users #177 and #315 also play an important role in linking the supervision company with other groups.

As mentioned above, the modularity of the network represents the possibility of dividing the network into different groups and communities. Therefore, modularity-based community detection was conducted using Gephi, and seven communities were identified. The left part of Figure 7 shows the seven communities detected with their silhouettes highlighted, and the nodes of the network are colored to represent different communities. Meanwhile, the top 16 users selected by betweenness are also marked with their IDs in Figure 7. The results showed that 15 of 16 participants were in the central part of the network with the exception of user #315. Among the top 16 lists, users #255, #161, #275, #62, #65, and #289 all belong to community #3, which has the highest number of the top 16 users among the seven communities. There are five users (#12, #51, #177, #282, #149) from community #2. It can be seen from the right part of Figure 7 that all these users have a strong connection with nodes of the same group and other groups. For example, there are many edges between user #12 and users of community #3, and quite a few links also exist between nodes of community #2 and user #12. In this case, removing user #12 from the network would significantly affect the efficiency of information sharing between communities #2 and #3. Following the same logic, users #285, and #198 also play an important role in bridging other participants and communities #5 and #6, respectively. Thus, more efforts should be made to detect brokers, as mistakes or problems originating from them significantly impact the overall efficiency of collaboration.

Finally, considering the three different centrality measures together, it is found that users #255, #62, and #65 are the three most important hubs and brokers in the network, as they appear in all the three top 16 user lists. As a result, managers should pay more attention to them, and determine a better way to further explore the efficiency of collaboration and information sharing.

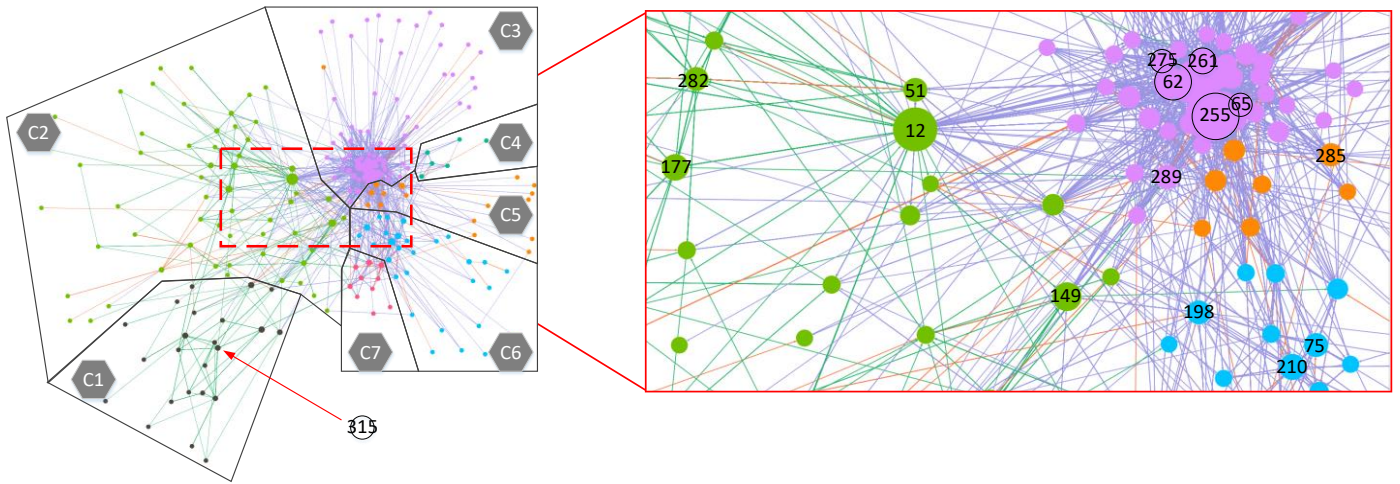

Figure 7 Detected Brokers and Communities

## 4.2 Identification of Frequently Collaborating Users

After important players such as hubs and brokers are detected and analyzed, this section further explores the connections between users by identifying frequently collaborating users based on ISF and LISF. If we take $ISF_{min}$ as 100, which means we only consider node pairs between which at least 100 data records are shared, Figure 8 is obtained. Similarly, following the method proposed in Section 3.3, a multiple graph as shown in Figure 9, is created by setting the value of $LISF_{min}$ as 60.

The left part of Figure 8 shows the generated network, and where a wider edge represents a larger ISF. According to Figure 8, the weights of the edges between node pairs (#210, #75), (#210, #264), (#231, #204), (#231, #199), (#255, #62), and (#255, #291) are quite high, implying that information is shared between them frequently and strong connections are established between these user pairs. In addition, the roles of frequently collaborating user pairs are owners and safety engineers. Furthermore, it seems that frequently collaborating users may be divided into three groups. User #255 is the center of group 1, and several users have strong relationships with user #255. The key players of group 2 are users #210, #75, and #264, the latter two of which are both strongly connected to user #210, and some other users are also linked to them. Group 3 also had three key players, they are users #231, #199, and #204.

According to the right part of Figure 9, frequently collaborating users can be divided into four groups: G1, G2, G3, G4. G1 is composed of user #513 and #310, between whom high-level issue records are usually transferred. G2 has the highest number of users, and the dominant level of on-site issues in this group is high. In G2, users #255 and #62 have the strongest connection in handing high-level issues. As for G3, all levels of on-site issues are involved; among them, low-level issues are usually handled by users #210, #75, and #264. Finally, G4 contains only four users and mainly deals with medium-level issues.

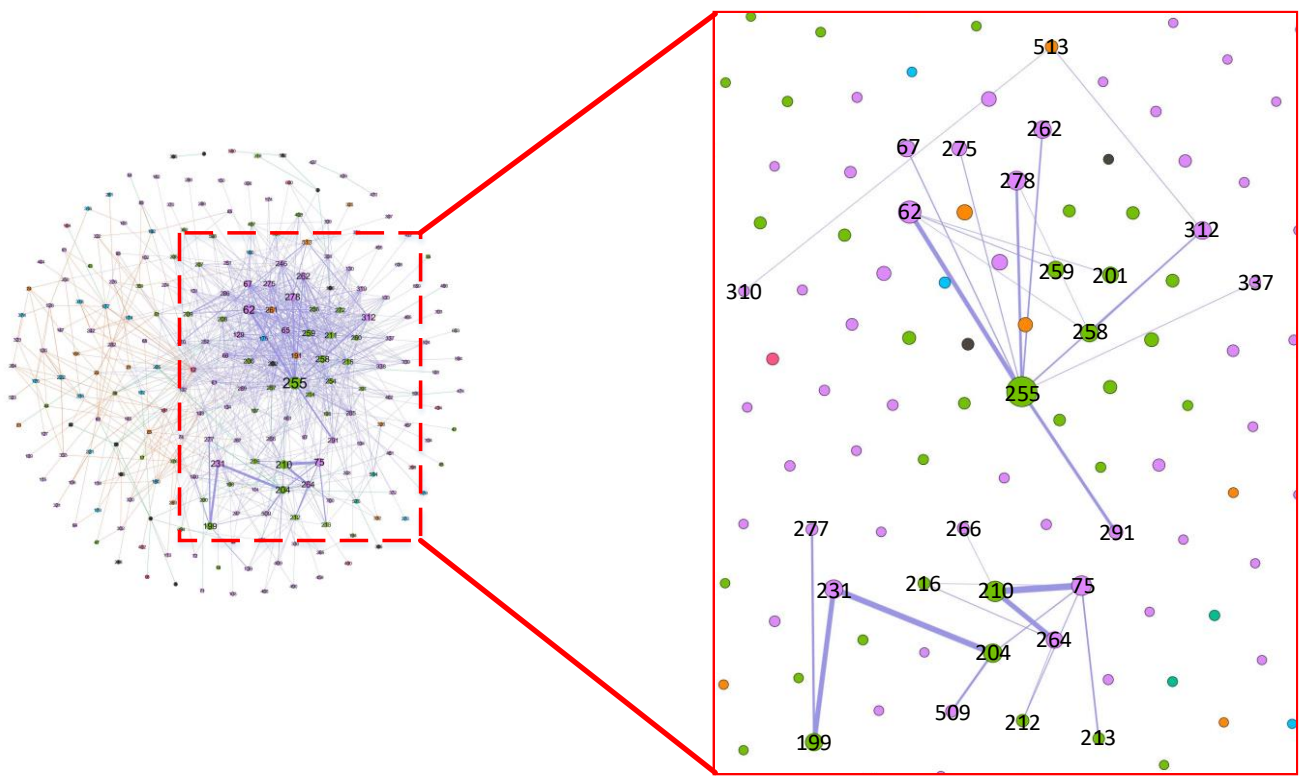

Figure 8 Frequently Collaborating Users Found with a Minimal ISF of 100

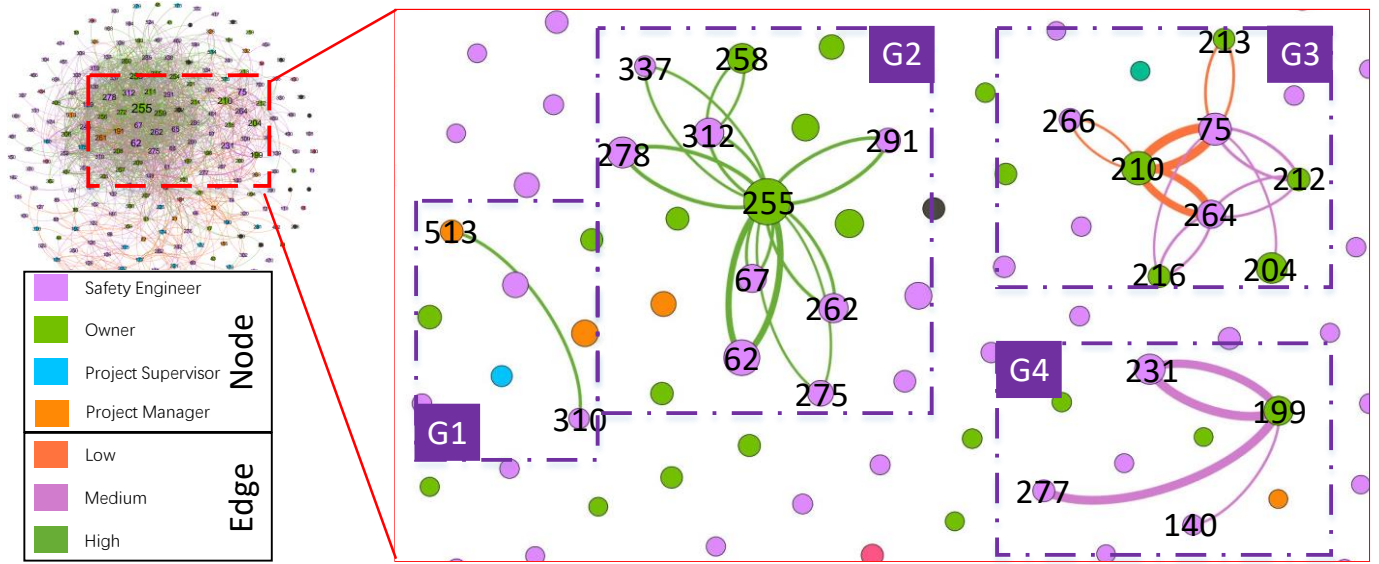

Figure 9 Frequently Collaborating Users Labeled by Task Levels

In this way, frequently collaborating users such as (#255, #62) and (#231, #199) could be identified, and the manner in which different user pairs are connected with task levels is also discovered. Thus, managers could deepen their understanding of the collaborative network, and make better decisions. For example, in the case of Figure 9, if a manager wants to know how high-level issues are handled, he/she should talk to users #255 and #62, whereas for medium-level tasks, #231, #199, and #277 are better candidates to provide suggestions.

### 4.3 Association Rule Mining of Information Flows

To explore associations between edges or information flows and task levels, the Apriori-based algorithm in Section 3.3 is adopted to discover hidden information flow patterns based on the parameters in Table 5.

Table 5 Parameters Used for Association Rule Mining

| Description | $S_{min}$ | $C_{min}$ | $L_{min}$ |
|---|---|---|---|
| Associations between Information Flows | 100/7250≈0.014 | 0.75 | 3 |
| Associations between Information Flows and Task Levels | 60/7250≈0.008 | 0.75 | 3 |

(1) Associations between Information Flows

Table 6 shows extracted association rules with minimal support $S_{min}$ as $100/7250 \approx 0.014$, which is obtained by dividing total number of issue records by 100, a minimal number of information flows happened between two users. Meanwhile, minimal confidence $C_{min}$, and minimal lift $L_{min}$ are chosen as 0.75 and 3 respectively. As illustrated in the first row of Table 6, if a data record goes from user #278 to #255 (noted as "278->255"), it is likely that the data record will go back from user #255 to #278 with a confidence as 0.9917. The lift of this pattern is 59.9, which is much higher than 1.0, indicating that there is a strong association between "278->255" and "255->278". In this way, it could be concluded that information is transferred forward and backward between node pairs (#255, #278), (#255, #291), (#255, #62), (#210, #264), (#210, #75), and (#231, #199) quite frequently. That is, for these frequently collaborating user pairs, we could expect that if an issue record is sent from one to another then the record will be sent back. This is consistent with the conclusion drawn above with respect to ISF. However, although ISF suggests that there is a strong relationship between users #231 and #204, no clear pattern is found between them. In other words, although many data records were shared between #231 and #204, they were then transferred to different users for further processing. Moreover, collaborative pattern that describes information sharing among three users, #199, #231, and #277, is also discovered, about which ISF provides no evidence.

Table 6 Frequent Collaborative Patterns

| No. | Patterns | Antecedent | Consequent | Support | Confidence | Lift |
|---|---|---|---|---|---|---|
| 1 | {'278->255', '255->278'} | {'278->255'} | {'255->278'} | 0.01656 | 0.9917 | 59.90 |
| 2 | {'278->255', '255->278'} | {'255->278'} | {'278->255'} | 0.01656 | 1.0000 | 59.90 |
| 3 | {'291->255', '255->291'} | {'291->255'} | {'255->291'} | 0.01876 | 0.9577 | 51.04 |
| 4 | {'291->255', '255->291'} | {'255->291'} | {'291->255'} | 0.01876 | 1.0000 | 51.04 |
| 5 | {'62->255', '255->62'} | {'62->255'} | {'255->62'} | 0.02497 | 1.0000 | 39.39 |
| 6 | {'62->255', '255->62'} | {'255->62'} | {'62->255'} | 0.02497 | 0.9837 | 39.39 |
| 7 | {'210->264', '264->210'} | {'210->264'} | {'264->210'} | 0.02608 | 0.9895 | 37.35 |
| 8 | {'210->264', '264->210'} | {'264->210'} | {'210->264'} | 0.02608 | 0.9844 | 37.35 |
| 9 | {'199->277', '199->231'} | {'199->277'} | {'199->231'} | 0.02856 | 1.0000 | 34.35 |
| 10 | {'199->277', '199->231'} | {'199->231'} | {'199->277'} | 0.02856 | 0.9810 | 34.35 |
| 11 | {'231->199', '199->277', '199->231'} | {'199->231'} | {'231->199', '199->277'} | 0.02621 | 0.9005 | 34.35 |
| 12 | {'231->199', '199->277', '199->231'} | {'231->199', '199->277'} | {'199->231'} | 0.02621 | 1.0000 | 34.35 |
| 13 | {'231->199', '199->277', '199->231'} | {'199->277'} | {'231->199', '199->231'} | 0.02621 | 0.9179 | 34.29 |
| 14 | {'231->199', '199->277', '199->231'} | {'231->199', '199->231'} | {'199->277'} | 0.02621 | 0.9794 | 34.29 |
| 15 | {'231->199', '199->231'} | {'231->199'} | {'199->231'} | 0.02677 | 0.9949 | 34.17 |
| 16 | {'231->199', '199->231'} | {'199->231'} | {'231->199'} | 0.02677 | 0.9194 | 34.17 |
| 17 | {'231->199', '199->277'} | {'231->199'} | {'199->277'} | 0.02621 | 0.9744 | 34.12 |
| 18 | {'231->199', '199->277'} | {'199->277'} | {'231->199'} | 0.02621 | 0.9179 | 34.12 |
| 19 | {'231->199', '199->277', '199->231'} | {'231->199'} | {'199->231', '199->277'} | 0.02621 | 0.9744 | 34.12 |

| 20 | {'231->199', '199->277', '199->231'} | {'199->277', '199->231'} | {'231->199'} | 0.02621 | 0.9179 | 34.12 |
| 21 | {'210->75', '75->210'} | {'75->210'} | {'210->75'} | 0.03091 | 1.0000 | 31.65 |
| 22 | {'210->75', '75->210'} | {'210->75'} | {'75->210'} | 0.03091 | 0.9782 | 31.65 |

As a result, beyond detected hubs or brokers, collaborative patterns including frequently collaborating users and associations between information-sharing paths could also be identified, enabling managers to understand the collaborative network more deeply.

(2) Associations between Information Flows and Task Levels

Meanwhile, if we consider the severity level of issue records as a label of edges, Table 7 is obtained based on the Apriori algorithm with minimal support $S_{min}$ as $60/7250 \approx 0.008$, minimal confidence $C_{min}$ as 0.75, and minimal lift $L_{min}$ as 3. According to Table 7, most of the high-level issues are related to user #255, and sent forward and backward between user #255 and #67, #312, #262, #291, #278, and #62, respectively. However, medium-level issues are processed by user pairs including (#255, #291), (#75, #212), (#199, #277), (#199, #231) and a triple (#199, #277, #231). In addition, low-level issues are handled by user pairs (#210, #264) and (#210, #75). Given that the lift of each pattern listed in Table 7 is much higher than 1.0, it is concluded that if an issue record with a specific level is sent from a user to another of the mentioned user pair, it is most likely that the issue record will be sent back with processing results.

Interestingly, according to Table 7, both high- and medium-level issues are frequently sent forward and backward between users #255 and #291, however, it can only be seen in Figure 9 that there is a strong association between high-level issues and user pair (#255, #291). This is because the Apriori algorithm provides more details about the associations between frequently collaborating users and task levels than LISF. Moreover, although the support of pattern No. 1 is lower than that of pattern No. 11, its lift is higher, indicating that the association between medium-level issues and user pair (#255, #291) is stronger than the association between high-level issues and user pair (#255, #291).

By identifying associations between task levels and frequently collaborating users, it is possible for managers to understand how well an engineer's experience or knowledge matches the severity levels of on-site tasks. Thus, managers could optimize the collaborative network and assign people to the positions where they are most needed.

Table 7 Frequent Collaborative Patterns Labeled by Task Level

| No. | Patterns | Antecedent | Consequent | Support | Confidence | Lift |
|---|---|---|---|---|---|---|
| 1 | {'291->255:M', '255->291:M'} | {'291->255:M'} | {'255->291:M'} | 0.00855 | 0.9841 | 115.05 |
| 2 | {'291->255:M', '255->291:M'} | {'255->291:M'} | {'291->255:M'} | 0.00855 | 1.0000 | 115.05 |
| 3 | {'255->67:H', '67->255:H'} | {'67->255:H'} | {'255->67:H'} | 0.00855 | 1.0000 | 113.25 |
| 4 | {'255->67:H', '67->255:H'} | {'255->67:H'} | {'67->255:H'} | 0.00855 | 0.9688 | 113.25 |
| 5 | {'75->212:M', '212->75:M'} | {'75->212:M'} | {'212->75:M'} | 0.00869 | 0.9692 | 111.51 |
| 6 | {'75->212:M', '212->75:M'} | {'212->75:M'} | {'75->212:M'} | 0.00869 | 1.0000 | 111.51 |
| 7 | {'255->312:H', '312->255:H'} | {'312->255:H'} | {'255->312:H'} | 0.00911 | 0.9851 | 106.56 |
| 8 | {'255->312:H', '312->255:H'} | {'255->312:H'} | {'312->255:H'} | 0.00911 | 0.9851 | 106.56 |
| 9 | {'262->255:H', '255->262:H'} | {'262->255:H'} | {'255->262:H'} | 0.00924 | 0.9853 | 103.50 |
| 10 | {'262->255:H', '255->262:H'} | {'255->262:H'} | {'262->255:H'} | 0.00924 | 0.9710 | 103.50 |
| 11 | {'291->255:H', '255->291:H'} | {'291->255:H'} | {'255->291:H'} | 0.01021 | 0.9367 | 91.75 |
| 12 | {'291->255:H', '255->291:H'} | {'255->291:H'} | {'291->255:H'} | 0.01021 | 1.0000 | 91.75 |
| 13 | {'278->255:H', '255->278:H'} | {'255->278:H'} | {'278->255:H'} | 0.01159 | 1.0000 | 85.27 |
| 14 | {'278->255:H', '255->278:H'} | {'278->255:H'} | {'255->278:H'} | 0.01159 | 0.9882 | 85.27 |
| 15 | {'62->255:H', '255->62:H'} | {'255->62:H'} | {'62->255:H'} | 0.01973 | 0.9795 | 49.64 |
| 16 | {'62->255:H', '255->62:H'} | {'62->255:H'} | {'255->62:H'} | 0.01973 | 1.0000 | 49.64 |
| 17 | {'210->264:L', '264->210:L'} | {'264->210:L'} | {'210->264:L'} | 0.02414 | 0.9887 | 40.72 |
| 18 | {'210->264:L', '264->210:L'} | {'210->264:L'} | {'264->210:L'} | 0.02414 | 0.9943 | 40.72 |
| 19 | {'210->75:L', '75->210:L'} | {'210->75:L'} | {'75->210:L'} | 0.02815 | 0.9808 | 34.85 |
| 20 | {'210->75:L', '75->210:L'} | {'75->210:L'} | {'210->75:L'} | 0.02815 | 1.0000 | 34.85 |
| 21 | {'199->277:M', '199->231:M'} | {'199->231:M'} | {'199->277:M'} | 0.02856 | 0.9810 | 34.35 |
| 22 | {'199->277:M', '199->231:M'} | {'199->277:M'} | {'199->231:M'} | 0.02856 | 1.0000 | 34.35 |
| 23 | {'199->277:M', '199->231:M', '231->199:M'} | {'199->231:M'} | {'199->277:M', '231->199:M'} | 0.02621 | 0.9005 | 34.35 |
| 24 | {'199->277:M', '199->231:M', '231->199:M'} | {'199->277:M', '231->199:M'} | {'199->231:M'} | 0.02621 | 1.0000 | 34.35 |
| 25 | {'199->277:M', '199->231:M', '231->199:M'} | {'199->277:M'} | {'231->199:M', '199->231:M'} | 0.02621 | 0.9179 | 34.29 |
| 26 | {'199->277:M', '199->231:M', '231->199:M'} | {'231->199:M', '199->231:M'} | {'199->277:M'} | 0.02621 | 0.9794 | 34.29 |
| 27 | {'231->199:M', '199->231:M'} | {'231->199:M'} | {'199->231:M'} | 0.02677 | 0.9949 | 34.17 |
| 28 | {'231->199:M', '199->231:M'} | {'199->231:M'} | {'231->199:M'} | 0.02677 | 0.9194 | 34.17 |
| 29 | {'199->277:M', '231->199:M'} | {'231->199:M'} | {'199->277:M'} | 0.02621 | 0.9744 | 34.12 |
| 30 | {'199->277:M', '231->199:M'} | {'199->277:M'} | {'231->199:M'} | 0.02621 | 0.9179 | 34.12 |
| 31 | {'199->277:M', '199->231:M', '231->199:M'} | {'231->199:M'} | {'199->277:M', '199->231:M'} | 0.02621 | 0.9744 | 34.12 |
| 32 | {'199->277:M', '199->231:M', '231->199:M'} | {'199->277:M', '199->231:M'} | {'231->199:M'} | 0.02621 | 0.9179 | 34.12 |

## 5. Discussion

In this research, an approach to twinning and mining collaborative networks was proposed and validated in a construction project. Because of limited time, only functions related to on-site inspection of existing systems were updated, and then the collaborative network was established and analyzed to discover hidden knowledge about collaboration. However, the proposed framework could be adopted to collect collaboration data and identify collaborative patterns when processing other construction tasks and even tasks from other fields. For this purpose, relevant systems should be updated to automatically capture data of information sharing, including creators, time, senders, receivers, and forward records, and no extra data are needed. Then, following the data preprocessing and SNA methods introduced in this work, hidden collaborative patterns that are valuable for decision-making purposes could be extracted. Moreover, although Gephi was used for graph visualization and calculation of centrality measures in this research, it is possible to perform the same work with Python-based packages such as NetworkX[39]. Hence, all the data preprocessing and mining procedures could be embedded as a Python-based web service[7], thereby achieving a fully automated data-driven decision-making process.

Centrality measures are commonly used to identify key players in a social network. In this research, DC is used to detect users who handle the most issues. In other words, the workload of these users is much higher than the others. Proper strategies that balance the workload between different users would improve the overall collaboration performance. Important users detected by CC are usually leaders or managers, who can easily spread information to different stakeholders. In this research, it was found that there are several users with similar closeness to others, which means that the network is robust in efficient information sharing, because it is easy to find a new user for information distribution if one is not able to transfer the required data in time. Brokers identified by betweenness are important in bridging different groups, and a network with fewer brokers is vulnerable with respect to information distribution and collaboration. This is because a small group of the network would lose their connections to others if the brokers are out of service. Occasionally, a certain user may be highlighted by multiple centrality measures, such as users #255, #62, and #65 in this research. If this is the case, more attention should be paid to them, because problems that hinder their functionality will cause significant problems to the overall network.

According to the defined ISF and LISF, and association rule mining utilized in this research, it was found that users tend to form small groups with two or three people for certain types or levels of issues. The information of an on-site issue is usually sent forward and backward between users of the small group, and it is also shown that there is a strong connection between frequently collaborating users and task levels. In this way, users are concentrated on certain levels of issues and collaborating with partners they are familiar with, thus achieving more efficient collaboration. Moreover, ISF and LISF could identify most of the frequently collaborating users and their connections with task levels, whereas association rule mining goes deeper in finding associations between information flows and task levels than ISF and LISF. Given that ISF and LISF are much easier and easy to implement, they are recommended for fast decision-making processes.

Owing to a lack of detailed information of the collaborative network, previous work related to SNA in the construction field mainly focused on the overall features of a network, and the identification of key players. This research attempts to take a step forward in discovering hidden knowledge from the connections between different stakeholders, and extracting associations between information flows and task levels. With the method proposed in this work, a much deeper understanding of collaborative networks could be achieved, thereby leading to a better decision-making process.

## 6. Conclusion and Future Work

In this research, an approach to twinning and mining collaborative networks is proposed and validated with collected collaborative data related to on-site inspection of a construction project. To fill the gap that most of collaborative networks are created manually, a strategy to update existing information systems and inject components for automatic data collection and network creation is introduced. Therefore, SNA and the Apriori algorithm are used to explore an established collaborative network in three aspects: 1) detection of hubs and brokers, which discovers users handling most of the on-site issues, with high capacity of information distribution or bridging different groups; 2) identifying frequently collaborating users and their relations to task levels; 3) unveiling associations between information flows and task levels. The benefits of the proposed approach are to equip managers with methods for evaluating the performance of collaboration, detection of key players, and understanding how people collaboratively complete construction tasks.

The contributions of this work to the body of knowledge are twofold: 1) metrics including ISF and LISF are defined and the Apriori algorithm is proposed for identification of collaborative patterns, i.e., frequently collaborating users and information flow patterns; 2) an integrated framework for automatic twinning and mining of fine-grained collaborative networks is established. Comparing to previous works, this research introduces methods and tools to look at collaboration in a much deeper way. Beyond detection of key players and network-level assessment of collaboration, strength of connections and association rules between information flows are investigated, making it possible to identify frequently collaborating users and unveil information flow patterns. In this way, insights on how people are collaborating with each other and how information is flowing from one another could be obtained, and managers can make wise decisions to make the collaboration and communication process more fluent and efficient.

Meanwhile, this work also contributes to the practice by:1) revealing that people tend to form small groups (pairs or threesomes in this research) to handle certain types or levels of tasks more efficiently; 2) providing a few easy-to-use metrics to discover collaborative patterns for decision-making purposes. In other words, instead of large, complex connected groups, construction managers should setup small groups with two or three works to handle on-issues in a more efficient way. And they could find frequently collaborating users easily with ISF or LISF to check if large or complex connected groups exists.

However, mining collaborative network in the construction domain is still in its infancy, and many investigations are needed. For example, relationships between topics of on-site issues[7] and frequently collaborating users, and the influence of modularity of the collaborative network on the performance of information flow, are two interesting topics worth exploring. Comparison of the proposed approach in this work with new methods introduced in the future is also value and important. Furthermore, integration with semantic analysis and reasoning[40,41], etc., are also encouraged to discover connections between spaces, objects, workers, tasks, and events for data-driven decision-making.

## 7. Acknowledgments

This research is supported by the National Natural Science Foundation of China (No. 51908323), the National Key R&D Program of China (No. 2018YFD1100900), the Tsinghua University Initiative Scientific Research Program (No. 2019Z02UOT) and the Beijing Natural Science Foundation (No. 8194067).